\documentclass{article}
\usepackage{graphicx}
\usepackage{amssymb}
\begin{document}

\title{Efficient Quantum Algorithms\\
       for State Measurement and Linear Algebra Applications}
\author{Apoorva Patel$^*$ and Anjani Priyadarsini$^\dagger$\\
        Centre for High Energy Physics, Indian Institute of Science,\\
        Bangalore 560012, Karnataka, India\\
        $^*$adpatel@iisc.ac.in\\
        $^\dagger$anjanipriyav@iisc.ac.in\\}

\date{}
\maketitle

\begin{abstract}
We present an algorithm for measurement of $k$-local operators in a quantum
state, which scales logarithmically both in the system size and the output
accuracy. The key ingredients of the algorithm are a digital representation
of the quantum state, and a decomposition of the measurement operator in
a basis of operators with known discrete spectra. We then show how this
algorithm can be combined with (a) Hamiltonian evolution to make quantum
simulations efficient, (b) the Newton-Raphson method based solution of
matrix inverse to efficiently solve linear simultaneous equations, and
(c) Chebyshev expansion of matrix exponentials to efficiently evaluate
thermal expectation values. The general strategy may be useful in solving
many other linear algebra problems efficiently.
\end{abstract}


\section{Introduction}

Simulations of quantum systems on classical computers are hard, but they
are expected to become easy on quantum computers by converting their
parallel implementations to superpositions---that was pointed out by Feynman
as a major motivation for developing quantum computers \cite{feynman}.
Such simulations would model physical systems directly into the quantum
hardware, but with greater freedom in the choice of parameters than the
limited values the natural systems have. That can be of great help in
understanding their dynamical properties, and this is a likely area where
quantum computers will demonstrate their superiority over classical ones
in near future.

Simulation problems are function evaluation problems. Their computational
complexity has to be measured in terms of both the input size and the
output precision. Efficient algorithms are those with the computational
complexity polynomial in the input as well as the output number of bits.
Conventional computational complexity analysis focuses on input size
dependence of decision problems (with just one output bit), and study
of output efficient algorithms is relegated to design of optimal methods
for arbitrary precision numerical analysis \cite{brent}. Optimisation of
output precision dependence of quantum simulation problems has attracted
attention relatively recently \cite{BCCKS1,BCCKS2,qhamevol}. In this
context, we have defined the computational complexity class P:P, which
is the set of computational problems that can be solved with resources
polynomial in the input size as well as the output size \cite{qhamevol}.
In the present work, we demonstrate that several practical quantum
linear algebra problems belong to this class.

Many simulation problems can be expressed as time evolution under specified
interactions, from some simple initial state to the final state whose
properties are to be determined. In this setting, the Hamiltonian evolution
problem has been extensively investigated. Consider a many-body quantum
system. Quantum simulations can sum multiple evolutionary paths contributing
to a quantum process in superposition at one go, while classical simulations
need to evaluate these paths one by one. Real physical systems are often
governed by local Hamiltonians, i.e. where each component interacts with
only a limited number of its neighbours, independent of the overall size
of the system. Early quantum evolution algorithms exploited this property
for efficient use of time and space resources \cite{lloyd,aharonov}.
More recently, the error complexity of the evolution has been reduced from
power-law to logarithmic in the inverse error, using algorithms with large
time evolution steps \cite{BCCKS1,BCCKS2,qhamevol}.

Classical simulation algorithms find it convenient to separate the evolution
and the measurement parts. That is logical, since all properties of the final
state are accessible in classical computation. But the situation is different
for quantum simulations; the solution of any quantum simulation problem is
incomplete without a procedure to measure the desired final state observables.
Explicitly, in the $2^n$-dimensional Hilbert space of $n$ qubits, we can
superpose $2^n$ components evolving in parallel, but we can measure only $n$
binary observables at the end. So the exponential gain of superposition is
limited by the restriction to extract only a small number of results at the
end. This dichotomy means that quantum algorithms will be advantageous only
when the final observables are local in some manner, and no general
prescription is available.

The problem of how to efficiently measure expectation values of final state
observables, after the time evolution, has not been adequately investigated
in the existing literature. Since quantum measurements are inherently
probabilistic, determination of the expectation values needs multiple
repetitions of the same algorithm. Thereafter, importance sampling or phase
estimation based methods yield errors that decrease as power-laws in the
number of repetitions \cite{harrow,clader}, and that is not efficient.
What we want is a strategy that decreases the errors exponentially with the
number of repetitions. While that is not possible for generic observables,
it can be achieved for $k$-local observables that appear in evaluations of
$k$-point Green's functions for many-body systems. In what follows, we
explicitly show how to do that.

To construct our algorithms, we first look at well-known output efficient 
classical algorithms, and select from them those that can be maximally
parallelised. Then we convert the parallel evaluation structure to quantum
superposition, to make the algorithms input efficient as well. In this
strategy, the mathematical techniques used in various components of the
algorithms are familiar; we just put them together in a clever way. It is
worth noting that the best classical algorithms may not offer the best
parallelisation that can be exploited by quantum superposition. In such
situations, our best quantum algorithms are related to the maximally
parallelisable classical algorithms, and are unrelated to the best
classical algorithms.

We first explain our digital quantum computation framework in Section
\ref{digitalqc}, expanding on our previous presentation \cite{qhamevol}.
Then, we describe our efficient measurement procedure in Section
\ref{effmeas}, in the context of quantum simulations. Afterwards,
in Sections \ref{effmatinv} and \ref{effmatexp} respectively, we
combine it with an algorithm for matrix inverse to solve simultaneous
linear equations, and an algorithm for matrix exponentiation to evaluate
thermal expectation values. In all cases, we compare the computational
complexity of our results with known classical and quantum algorithms,
and point out the improvements made. Our results for the computational
complexity of measuring $k$-point Green's functions, of evolving a
quantum state with the given Hamiltonian, of solving linear simultaneous
equations and of exponentiating a sparse matrix, are contained in
Eqs.(\ref{cmplx_meas},\ref{cmplx_hamevol},\ref{cmplx_NR},\ref{cmplx_exp})
respectively.

\section{Digital Quantum Computation}
\label{digitalqc}

It is routine to represent a quantum state in an $N$-dimensional Hilbert
space as
\begin{equation}
|x\rangle = \sum_{j=0}^{N-1} x_j |j\rangle ~,~~
\sum_{j=0}^{N-1} |x_j|^2 = 1 ~,
\end{equation}
where $x_j$ are continuous complex variables. This analog representation
is not convenient for high precision calculations, because any physical
apparatus can determine continuous values up to only a limited precision.
A digital representation bypasses this limitation, by breaking up $x_j$
into a sequence of digits, where each digit has only a finite number of
possibilities that can be easily distinguished, and the number of digits
can be made as large as desired. Of course, the benefit of digitisation
is maximised when the complete calculation, from the input to the output,
is carried out in the digital representation. We describe here the
ingredients needed to execute a quantum computation in such a digital mode.

\smallskip
\noindent{\bf States:}
We use the digital representation specified by the map \cite{qhamevol}:
\begin{equation}
\label{digitalrep}
|x\rangle \rightarrow \frac{1}{\sqrt{N}}
\sum_{j=0}^{N-1} |j\rangle |x_j\rangle_q ~,
\end{equation}
which mimics the storage of a vector in classical computer registers.
It is a quantum state in a $(2^q N)$-dimensional Hilbert space, where
$|x_j\rangle_q$ are the basis vectors of a $q$-bit register representing
the truncated values of $x_j$ (a complex number $x_j$ can be represented
by a pair of real numbers, and all $2^q x_j$ are truncated to integers).
This representation is fully entangled between the component index state
$|j\rangle$ and the register value state $|x_j\rangle_q$, with a unique
non-vanishing $|x_j\rangle_q$ (out of $2^q$ possibilities) for every
$|j\rangle$. It is important to observe that no constraint is necessary
on the register values $x_j$ in this representation---the algorithm has
to take care of the overall unitary evolution. This freedom allows simple
implementation of all linear algebra operations (and not just unitary
transformations) on $|x_j\rangle_q$, transforming them among the $2^q$
basis states using only C-not and Toffoli gates of classical reversible
logic, with the index state $|j\rangle$ acting as control. For example,
\begin{eqnarray}
\label{lindigitA}
c|x\rangle &\rightarrow&
\frac{1}{\sqrt{N}} \sum_{j=0}^{N-1} |j\rangle |cx_j\rangle_q ~, \\
|x\rangle + |y\rangle &\rightarrow&
\frac{1}{\sqrt{N}} \sum_{j=0}^{N-1} |j\rangle |x_j+y_j\rangle_q ~,
\label{lindigitB}
\end{eqnarray}
map non-unitary operations on the left to unitary operations on the right.
These elementary operations can be combined to construct any power series.%
\footnote{This is a much simpler procedure than the ``addition of unitaries"
used in Ref.~\cite{BCCKS2}.}
Note that a crucial requirement for implementing linear algebra operations
in the digital representation is that only a single index (``$j$" in the
preceding formulae) controls the whole entangled state.

\smallskip
\noindent{\bf Observables:}
The freedom to choose a convenient representation for the quantum states
is particularly useful due to the fact that the quantum states are never
physically observed. All physically observed quantities are the expectation
values of the form $\langle x|O_a|x \rangle$. So the digital representation
is completed by constructing for each observable $O_a$ in the $N$-dimensional
Hilbert space a related observable $\tilde{O}_a$ in the $(2^q N)$-dimensional
Hilbert space, such that
\begin{equation}
\label{Odigitexp}
\langle x| O_a |x \rangle
  = \sum_{j,l=0}^{N-1} x_j^* x_l \langle j| O_a |l \rangle
  = \frac{1}{N}\sum_{j,l=0}^{N-1}
      {}_q\langle x_j| \langle j| \tilde{O}_a |l \rangle |x_l \rangle_q ~.
\end{equation}
For this equality to hold, it suffices to construct the operator
$\tilde{O}_a = O_a \otimes O_q$, where the Hermitian operator $O_q$
in the $2^q$-dimensional Hilbert space satisfies
\begin{equation}
\label{digitalop}
\langle x_j | O_q |x_l \rangle = N x_j^* x_l ~.
\end{equation}
To this end, we note that
\begin{equation}
\langle x_j| (1+\sigma_1)^{\otimes q} |x_l \rangle = 1 ~,
\end{equation}
and the place-value operator for a bit-string,
\begin{equation}
\label{plvalop}
V = \sum_{k=0}^{q-1} 2^{-k} I^{\otimes k} \otimes
    \Big(\frac{1-\sigma_3}{2}\Big) \otimes I^{\otimes (q-k-1)} ~,
\end{equation}
gives $V|x_j\rangle = x_j|x_j\rangle$. The solution to Eq.(\ref{digitalop}),
therefore, is independent of the quantum state and the observable:%
\footnote{In terms of the uniform superposition state
          $|s\rangle_q=H^{\otimes q}|0\rangle_q$,
          $(1+\sigma_1)^{\otimes q} = 2^q|s\rangle_q~{}_q\langle s|$.
          Also, the factor $(1+\sigma_1)^{\otimes q}$ can be omitted
          from $O_q$ in evaluation of $\langle x_j|O_q|x_j\rangle$.}
\begin{equation}
\label{digitop}
O_q = N V^\dagger (1+\sigma_1)^{\otimes q} V ~.
\end{equation}
Since $V$ is a sum of $q$ bit-wise fully factorised terms, $O_q$ can be
expressed as a sum of $q^2$ such terms, and the computational complexity
of measurement of physical observables in the digital representation is
$O(q^2)$ times that in the analog representation. Moreover, the bit-wise
separated structure of $V$ allows evaluation of any single specific bit
of $\langle x|O_a|x \rangle$, if so desired, with $O(q)$ extra effort
compared to the analog case. More generally, any function $f(x_j)$ for
the state $|x_j\rangle$ can be computed using just the machinery of
classical reversible logic, and overall normalisations can be adjusted
at the end of the calculation.

We essentially bypass the constraint of unitarity in the digital
representation, by using two different metrics in the $2^q$-dimensional
space of the coordinates $\{x_j\}$. The Cartesian metric is used for
implementing the linear algebra operations, and the metric $O_q$ is used 
for evaluating the expectation values of observables. This trick allows us
to exploit the advantages of the digital representation over the analog one,
i.e. easy implementation of arbitrary precision calculations and simple
linear algebra operations. It is worthwhile to note that construction of
fault-tolerant operations in the digital representation is considerably
simpler than in the analog case, because only a small set of quantum
logic gates is required---the C-not and the Toffoli gates to implement
Eqs.(\ref{lindigitA},\ref{lindigitB}) and the Hadamard gate to implement
Eq.(\ref{initstate}).
% Are there any operations (e.g. Hadamard) that become more complicated?

\smallskip
\noindent{\bf Initialisation:}
To efficiently incorporate the digital representation in an algorithm,
methods must be found to not only manipulate the register values
$|x_j\rangle$ efficiently, but also to initialise and to observe them.
At the start of the calculation, we need to assume that the initial values
$x_j(0)$ can be efficiently computed from $j$, say using the control
operation $C_x$. Then, for $N=2^n$, the initial state is created easily
using the Hadamard and the $C_x$ operations as
\begin{eqnarray}
\label{initstate}
|0\rangle |0\rangle_q
&& \mathop{\longrightarrow}\limits^{H^{\otimes n}\otimes I}~
\frac{1}{\sqrt{N}} \sum_{j=0}^{N-1} |j\rangle |0\rangle_q \\
&& \mathop{\longrightarrow}\limits^{C_x}~
\frac{1}{\sqrt{N}} \sum_{j=0}^{N-1} |j\rangle |x_j(0)\rangle_q ~.
\end{eqnarray}
When $N$ is not a power of 2, a simple fix is to enlarge the $j$-register
to the closest power of 2 and initialise the additional $x_j$ to zero.
Thereafter, the linear algebra operations can be implemented such that
the additional $x_j$ remain zero, and the overall normalisation (i.e.
$1/\sqrt{N}$) can be corrected in the final result as a proportionality
constant.

\smallskip
\noindent{\bf Evolution:} The evaluation of $|x(T)\rangle=U|x(0)\rangle$ for
a unitary operator $U$ is a matrix-vector product. That can be efficiently
calculated in the digital representation, when $U$ can be expressed as a
sum (or product) of a finite number of block-diagonal terms, e.g. using a
series expansion. The linear algebra operations can then combine all the
terms easily. It is not necessary for the intermediate steps involving
individual terms to satisfy the unitary constraint; it is sufficient that
the final result obeys $\sum_{j=0}^{N-1}|x_j(T)|^2=1$.

\smallskip
\noindent{\bf Measurement:}
At the end of the calculation, we need to assume that the final state
observables are efficiently computable from $x_j(T)$. In the analog
representation, determination of $|x_j(T)|^2$ is probabilistic, and
requires an ensemble of measurements covering the full range of the
index $j$. In contrast, in the digital representation, when a particular
$j$ is observed, the corresponding $x_j(T)$ can be determined exactly.
Then, the advantage of the digital representation is that the index $j$
can be handled in parallel (classically) or in superposition (quantum
mechanically). The fact that $V$ is an eigenoperator for $|x_j\rangle$
allows $\langle x_j|O_q|x_l\rangle$ to be evaluated deterministically
with $O(q^2)$ effort. That makes the efficiently measurable observables
those for which the sum over $N^2$ terms in Eq.(\ref{Odigitexp}) can be
evaluated with poly($n$) effort.

\smallskip
\noindent{\bf Density matrix:}
We point out that a digital representation for the density matrix can be
constructed in a completely analogous manner. The map for
\begin{equation}
\rho = \sum_{i,j=0}^{N-1} \rho_{ij} |j\rangle\langle i| ~,~~
\sum_{i=0}^{N-1} \rho_{ii} = 1 ~,
\end{equation}
describing pure as well as mixed states, is
\begin{equation}
\rho \rightarrow
  \frac{1}{N} \sum_{i,j=0}^{N-1} |j\rangle\langle i| ~ |\rho_{ij}\rangle_q ~,
\end{equation}
where $|\rho_{ij}\rangle_q$ are the basis vectors of a $2^q$-dimensional
Hilbert space representing the truncated values of $\rho_{ij}$. The most
general evolution of the density matrix is a completely positive
trace-preserving linear transformation, specified by a Kraus representation
\begin{equation}
\rho \rightarrow \sum_\mu M_\mu \rho M_\mu^\dagger ~,~~
\sum_\mu M_\mu^\dagger M_\mu = 1 ~.
\end{equation}
It is straightforward to implement that with operations similar to 
Eqs.(\ref{lindigitA},\ref{lindigitB}). Furthermore, expectation value
of any physical observable can be obtained as
\begin{equation}
Tr(\rho O_a) = \sum_{i,j=0}^{N-1} \rho_{ij} \langle i|O_a|j\rangle
             = 2^{q/2} \sum_{i,j=0}^{N-1} \langle i|O_a|j\rangle
             ~ {}_q\langle s|V|\rho_{ij}\rangle_q ~.
\end{equation}
Thus any single specific bit of $Tr(\rho O_a)$ can be evaluated with
the same effort as in the analog case, and the computational complexity
of measurement of an observable in the digital representation is $O(q)$
times that in the analog representation.

\smallskip
\noindent{\bf Computational complexity:}
Finally, we have to take care of the fact that a digital computation with
finite register size produces round-off errors, because real values are
replaced by integer approximations. With $q$-bit registers, the available
precision is $\delta=2^{-q}$. Using simple-minded counting, elementary
bit-level computational resources required for additions, multiplications
and polynomial evaluations are $O(q)$, $O(q^2)$ and $O(q^3)$ respectively.
(Overflow/underflow limit the degree of the polynomial to be at most $q$.)
Since all efficiently computable functions can be approximated by accurate
polynomials, the effort needed to evaluate individual elements of any
operator is thus $O(q^3)$.

Linear algebra algorithms are often dominated by operator-state products.
For $d$-sparse operators, their classical computational complexity is
$O(dNq^3)$. Such operators can be expressed as a sum of $d$ block-diagonal
operators with fixed block sizes, and the number of blocks is $O(N)$. When
an efficient labeling scheme for the blocks exists, the index $j$ can be
broken down into $O(n)$ tensor product factors (analogous to
Eq.(\ref{initstate})), and then quantum superposition makes the cost of
multiplying the operator with a state proportional to $n$. That makes the
quantum computational complexity of the operator-state product $O(dnq^3)$.

The register size $q$ is determined using the constraint that the round-off
error accumulated over the whole algorithm should not exceed the specified
error bound $\epsilon$. For an algorithm containing $r$ sparse operator-state
products, that can be achieved by choosing $dr\delta = O(\epsilon)$, which
gives $q=\Omega(\log(dr/\epsilon))$. This relation between $q$ and $\epsilon$
is crucial in construction of algorithms that are efficient with respect to
the output precision; a computational complexity of the form $poly(q)$
becomes $poly(\log\epsilon)$.

\section{Efficient Measurements}
\label{effmeas}

Hamiltonian simulation evolves an initial quantum state $|\psi(0)\rangle$
to a final quantum state $|\psi(T)\rangle$, in presence of interactions
specified by a Hamiltonian $H(t)$:
\begin{equation}
\label{unitevol}
|\psi(T)\rangle = U(T) |\psi(0)\rangle ~,~~
U(T) =  {\cal P} \Big[ \exp\big(-i\int_0^T H(t)~dt\big) \Big] ~.
\end{equation}
The initial state is usually easy to prepare, while the final state is
generally unknown. The path ordering of the unitary evolution operator
$U(T)$, denoted by the symbol ${\cal P}$ in Eq.(\ref{unitevol}), is needed
when various terms in the Hamiltonian do not commute. Properties of the
final state are subsequently extracted from expectation values of observables:
\begin{equation}
\label{expectobs}
\langle O \rangle =  \langle\psi(T)|O|\psi(T)\rangle ~.
\end{equation}
Efficient algorithms to determine the final state $\psi(T)$, up to a
specified error bound $\epsilon$ and for a certain class of Hamiltonians,
have been constructed in earlier works \cite{BCCKS1,BCCKS2,qhamevol}.
They have computational complexity
\begin{equation}
\label{cmplx_hamevol}
O\left( T\frac{\log(T/\epsilon)}{\log(\log(T/\epsilon))}\mathcal{C} \right),
\end{equation}
where $\mathcal{C}=O(dnq^3)$ is the computational complexity of evaluating
a sparse Hamiltonian-state product. Here, assuming that $\psi(T)$ is available,
we formulate a method to determine the expectation values $\langle O \rangle$
efficiently, up to a given precision $\epsilon$ and for a certain class of
observables. Together, they make a variety of quantum simulation problems
belong to the computational complexity class P:P. 

We concern ourselves here only with bounded operators,%
\footnote{Physical problems with unbounded operators exist, but their
          numerical solutions require more sophisticated techniques.}
acting in finite $N$-dimensional Hilbert spaces. A general operator would
then be a dense $N \times N$ matrix, and there is no efficient way to even
write it down. So we only look at operators with the following properties,
often encountered in physical problems:\\
(1) The Hilbert space is a tensor product of many small components of fixed
size, e.g. $N=2^n$ for a system of $n$ qubits. Generically, $|\psi(T)\rangle$
is an entangled state in this space.\\
(2) The operator is a tensor product of a finite number of local variables,
e.g. $O=\prod_{i=1}^k O_i$ where each $O_i$ is a single qubit observable at
a distinct location. More generally, each $O_i$ can be spread over a fixed
number of neighbouring components. Such operators appear in evaluations of
$k$-point Green's functions in many-body systems, and we call them $k$-local.\\
(3) The decomposition of each $O_i$ in terms of its elementary components
is efficiently computable, e.g. $O_i$ for a qubit is a linear combination
of Pauli operators with specific coefficients (which may depend on $i$).\\
These features allow a compact description of the observable $O$, and then
the resources required to just write it down do not influence determination
of its expectation value. Furthermore, evaluation of the sparse matrix-vector
product $O|\psi(T)\rangle$ can be easily parallelised, if necessary.

In terms of these specifications, \emph{operator expectation values for
problems in class P:P can be calculated using computational resources that
are polynomial in $\log(N)$ and $\log(\epsilon)$, with finite $k$}.

\subsection{Operator Decomposition}
\label{opdecomp}

Our efficient measurement strategy has two important ingredients.
The first ingredient is to decompose the operator as a sum of tensor
products of Pauli operators. Any single qubit $O_i$ can be expressed
in the Pauli basis as $(a_0I+a_1\sigma_x+a_2\sigma_y+a_3\sigma_z)_i$.
In this basis, we then have the decomposition:
\begin{equation}
\label{Paulidecomp}
O = \prod_{i=1}^k O_i = \sum_{j=1}^K \beta_j \Sigma_j ~,
\end{equation}
where each $\Sigma_j$ is a tensor product of $k$ Pauli operators at
different locations, with $K\le3^k$. Also, a Hermitian $O$ implies that
$\beta_j$ are real. When $O_i$ is spread over a finite cluster of $s$
qubits, its Pauli basis decomposition has $4^s$ terms, and the corresponding
Eq.(\ref{Paulidecomp}) has at most $K=4^{ks}$ terms. The important point is
that $K$ is finite when $k$ is finite.

In a more rigorous notation, one has to write a tensor product factor $I$
at each of the $n-k$ locations not covered by $O_i$. When calculating the
expectation values, these $n-k$ locations get summed over, resulting in a
reduced density matrix for the $k$ locations of $O_i$, and the non-trivial
part of measurement depends on this reduced density matrix only. For the
sake of simplicity, we avoid such an elaborate notation.

With our assumed properties of $O$, $K$ is a finite number and the
coefficients $\beta_j$ are easily computable (henceforth we take them
as known). The advantage of this decomposition are:\\
(a) All $\Sigma_j$ have well-separated eigenvalues $\pm1$ only, and their
eigenvectors are known. That simplifies their measurement \cite{CKWYZ}.\\
(b) The tensor product factors are decoupled, and so can be evaluated
in parallel. The value of $\Sigma_j$ can be accumulated in a single
ancilla qubit with $O(k)$ operations, which can then be determined by a
single binary measurement operation.%
\footnote{This is a far superior strategy compared to performing $k$
          binary measurements.}
This is completely analogous to the syndrome extraction procedure for
quantum error correction codes \cite{nielsen}.

The linear norm of $O$ is, $\|O\| = \sum_j |\beta_j| \cdot \|\Sigma_j\|
= \sum_j |\beta_j|$. Various $\Sigma_j$ may not commute with each other,
and so the $\langle\Sigma_j\rangle$ need to be evaluated one by one.
When each $\langle\Sigma_j\rangle$ is determined up to an additive error
$\epsilon$, the expectation value $\langle O\rangle$ is determined up to
an accuracy
\begin{equation}
\label{Oacc}
\epsilon_{\langle O\rangle} \le \sum_j |\beta_j|\epsilon = \epsilon \|O\| ~.
\end{equation}
With these definitions, the total computational complexity for determination
of $\langle O\rangle$ to fractional error $\epsilon$ is $O(Kkm)$, where $m$
is the number of measurement trials needed to determine each $\Sigma_j$
to additive error $\epsilon$.

\subsection{Measurement Optimisation}

\begin{figure}[t]
{
\begin{center}
%\setlength{\unitlength}{1mm}
% Standard unit length is 1pt
\begin{picture}(400,80)
\put(20,60){\line(1,0){40}}
\put(20,40){\line(1,0){18}}
\put(42,40){\line(1,0){18}}
\put(20,20){\line(1,0){40}}

\put(-4,48){$|\psi\rangle$}
\put(12,48){$\bigg\lbrace$}
\put(0,18){$|0\rangle$}

\put(40,60){\circle*{4}}
\put(40,40){\circle{4}}
\put(40,20){\circle{8}}
\put(40,16){\line(0,1){22}}
\put(40,42){\line(0,1){18}}

\put(80,18){\line(1,0){2}}
\put(80,62){\line(1,0){2}}
\put(82,18){\line(0,1){44}}
\put(86,48){$P_1^{(1)}P_0^{(2)}|\psi\rangle|1\rangle$}
\put(86,32){$+(1-P_1^{(1)}P_0^{(2)})|\psi\rangle|0\rangle$}
\put(60,14){\framebox(12,12){$\nearrow$}}
\put(35,0){(a)}

\put(220,60){\line(1,0){14}}
\put(246,60){\line(1,0){28}}
\put(286,60){\line(1,0){14}}
\put(220,40){\line(1,0){80}}
\put(220,20){\line(1,0){80}}

\put(196,48){$|\psi\rangle$}
\put(212,48){$\bigg\lbrace$}
\put(200,18){$|0\rangle$}

\put(234,54){\framebox(12,12){$H$}}
\put(274,54){\framebox(12,12){$H$}}
\put(260,60){\circle*{4}}
\put(280,40){\circle*{4}}
\put(260,20){\circle{8}}
\put(280,20){\circle{8}}
\put(260,16){\line(0,1){44}}
\put(280,16){\line(0,1){24}}

\put(320,18){\line(1,0){2}}
\put(320,62){\line(1,0){2}}
\put(322,18){\line(0,1){44}}
\put(326,48){$\frac{1}{2}(1-\sigma_x^{(1)}\sigma_z^{(2)})|\psi\rangle|1\rangle$}
\put(326,32){$+\frac{1}{2}(1+\sigma_x^{(1)}\sigma_z^{(2)})|\psi\rangle|0\rangle$}
\put(300,14){\framebox(12,12){$\nearrow$}}
\put(255,0){(b)}

\end{picture}
\end{center}
}
\caption{Quantum logic circuits for measuring (a) a product of two projection
operators, and (b) a product of two Pauli operators. \fbox{$\nearrow$} denotes
the binary measurement operation.}
\label{prodopmeasur}
\end{figure}
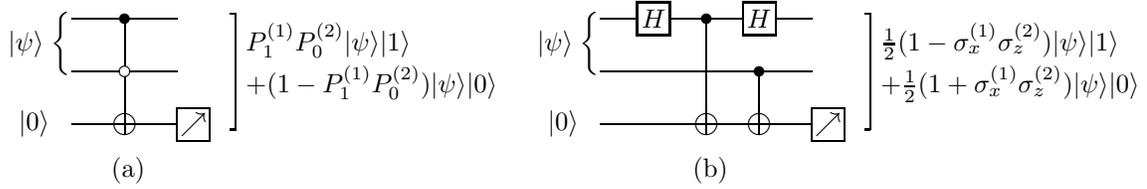

Consider evaluation of a single $\langle \Sigma_j \rangle$ in quantum theory.
Quantum measurements are probabilistic, and one has to repeat the process
many times, with identically prepared states, to obtain the result with high
accuracy. Individual measurements of $\Sigma_j$ yield binary results. Let
$p,1-p$ be the probabilities that $\Sigma_j$ is measured to be $+1,-1$
respectively. Then
\begin{equation}
\langle \Sigma_j \rangle = 2p-1 ~,
\end{equation}
i.e. the expectation value of $(1+\Sigma_j)/2$ is the probability that
$\Sigma_j$ is measured to be $+1$. Logic circuits for measuring $\Sigma_j$
are easily constructed, based on the facts that $\Sigma_j$ is unitary and
$(1\pm\Sigma_j)/2$ are projection operators. We use the $\sigma_z$
eigenstates, $|0\rangle$ and $|1\rangle$, as the computational basis.
Then the $\sigma_x,\sigma_y$ eigenvalues can be extracted by measurements
in suitably rotated bases, using the identities
$\sigma_x = H \sigma_z H, \sigma_y=i\sigma_z\sigma_x$,
where $H$ is the Hadamard operator. It is straightforward to collect
contributions of different Pauli factors making up $\Sigma_j$ into a single
ancilla qubit using controlled operations. Finally, binary measurement of
the ancilla qubit in the computational basis yields the value for $\Sigma_j$,
+1 or -1. As a simple illustration, Figs.\ref{prodopmeasur}a,b show the
quantum logic circuits for measuring a product of two projection operators
and a product of two Pauli operators respectively.

We obtain the accuracy with which $\langle\Sigma_j\rangle$ can be determined
after $m$ binary measurements, according to the well-known Chernoff bound
\cite{chernoff}. Let $X_i$ be independent random variables that take values
+1 or -1, with probability $p$ and $1-p$ respectively. Then their mean over
$m$ determinations, $\overline{X}=\frac{1}{m}\sum_{i=1}^m X_i$, converges to
$\mu=2p-1$ as $m\rightarrow\infty$. With finite $m$, and $s>0$,
\begin{equation}
\label{Markovineq}
Prob(\overline{X}-\mu>\delta) = Prob(e^{s\sum_i X_i} > e^{sm(\mu+\delta)})
= \langle e^{s\sum_i X_i} \rangle e^{-sm(\mu+\delta)},
\end{equation}
by Markov's inequality. Evaluation of the expectation value gives
\begin{equation}
\langle e^{s\sum_i X_i} \rangle = \prod_{j=1}^m (pe^s + (1-p)e^{-s})
\le e^{-ms} e^{mp(e^{2s}-1)} ~.
\end{equation}
Optimisation of the bound in Eq.(\ref{Markovineq}), with respect to the
parameter $s$, gives $e^{2s}=1+\Omega$ where $\Omega=\frac{\delta}{2p}$.
The overall bound for the upper tail probability then becomes, with
$0\le\delta\le2(1-p)$,
\begin{equation}
Prob(\overline{X}-\mu>\delta) \le \left(
\frac{e^\Omega}{(1+\Omega)^{1+\Omega}} \right)^{mp} ~.
\end{equation}
A similar analysis for the lower tail probability gives, with
$0\le\delta\le2p$,
\begin{equation}
Prob(\overline{X}-\mu<-\delta) \le \left(
\frac{e^{-\Omega}}{(1-\Omega)^{1-\Omega}} \right)^{mp} ~.
\end{equation}

Using the inequalities,
\begin{equation}
\ln(1+\Omega) > \frac{\Omega}{1+\Omega/2} ~,~~
% This is a truncation of the continued fraction expansion.
(1-\Omega)\ln(1-\Omega) > -\Omega + \frac{\Omega^2}{2} ~,
\end{equation}
the combined failure probability satisfies:%
\footnote{For binary random variables taking values $1$ or $0$,
          the bound is the same with $\Omega=\delta/p$.}
\begin{equation}
Prob(|\overline{X}-\mu|>\delta) < 2 e^{-mp\Omega^2/(2+\Omega)} ~.
\end{equation}
To make this failure probability less than a specified value $\epsilon_m$,
we need
\begin{equation}
\label{trialbound}
m > \ln(2/\epsilon_m) \left( \frac{2+\Omega}{p\Omega^2} \right)
  = \ln(2/\epsilon_m) \left( \frac{8p+2\delta}{\delta^2} \right) ~.
\end{equation}
The $O(Kkm)$ computational complexity for measurement then has logarithmic
dependence on $\epsilon_m$ as desired. Note also that the total success
probability for determination of $\langle O \rangle$ exceeds
$(1-\epsilon_m)^K > 1-K\epsilon_m$, which is greater than $1/2$ for
$\epsilon_m<1/(2K)$.

There is a problem, however, when $\mu$ is a continuous variable. That
would require small values of $\delta$ for high accuracy, and the bound
in Eq.(\ref{trialbound}) behaves as $1/\delta^2$ as $\delta\rightarrow0$.
This power law scaling is certainly undesirable. A way out is to make the
possible values of $\mu$ discrete. Then $\delta$ can be given a finite
value, e.g. half the separation between the discrete values of $\mu$, and
one does not have to worry about how the total computational complexity
depends on it. Such a discretised measurement process can be constructed
using the digital representation of quantum states, and we turn to that
in the next subsection.

\subsection{Discrete Optimal Measurements} 

In our digital representation described in Section \ref{digitalqc},
\begin{equation}
\langle x|\Sigma_i|x\rangle = \frac{1}{N}\sum_{j,l=0}^{N-1}
\langle j|\Sigma_i|l\rangle \langle x_j| O_q |x_l\rangle ~.
\label{sigmaexpect}
\end{equation}
In this expression, the quantities $\langle j|\Sigma_i|l\rangle$ are
fixed constants that can be evaluated at the outset, independent of the
state $|x\rangle$. Given the tensor product structure of $\Sigma_i$,
several simplifications can be carried out:\\
(1) The eigenbasis of $\Sigma_i$ is known. Performing measurements in this
basis reduces the double sum on r.h.s. of Eq.(\ref{sigmaexpect}) to a single
one, i.e. $j=l$.\\
(2) $\langle j|\Sigma_i|j\rangle$ are products of $k$ nontrivial factors,
and so can be easily evaluated for all values of $j$.\\
(3) Since $\Sigma_i$ have eigenvalues $\pm 1$, and $\langle x|x\rangle=1$,
it is sufficient to evaluate contribution of one of the eigenvalues to
$\langle\Sigma_i\rangle$, which amounts to restricting the sum on the
r.h.s. of Eq.(\ref{sigmaexpect}) to a subset of terms.

When a particular value of $\Sigma_i$ is extracted into an ancilla qubit,
say as illustrated in Fig.1(b), and a binary measurement of the ancilla
qubit is performed, all the unmeasured qubits are automatically traced over.
This partial trace gives the sum of the corresponding terms on the r.h.s.
of Eq.(\ref{sigmaexpect}), adding up to $p$ or $1-p$. Next, as per
Eqs.(\ref{plvalop},\ref{digitop}), $O_q$ is a sum of $q^2$ projection
operators with place-value weights, and each of the $q^2$ terms can be
extracted using an ancilla qubit as illustrated in Fig.1(a). Moreover,
in the digital representation, $|x_j\rangle$ is an eigenstate of $V$, and
so the result in the ancilla qubit is deterministic (and not probabilistic).
Adding all the $q^2$ results with their place-value weights, which can be
done classically, gives the total result for $\langle x_j| O_q | x_j\rangle$.

Combining measurements of both $\Sigma_i$ and $O_q$, we obtain $p$ in a
digital representation. Since each bit of $p$ can only take the two discrete
values, $0$ or $1$, it can be determined with high confidence (i.e. small
$\epsilon_m$) and a coarse-grained measurement window (i.e. $\delta=1/2$),
in the notation of the previous subsection. The crucial advantage provided
by the digital representation is that a probabilistic estimate of $p$ is
replaced by a deterministic evaluation of $p$, which can be carried out
bit-by-bit to any desired accuracy.

In this evaluation strategy, we have broken up $O$ as a sum of $O(Kq^2)$
discrete operators. Measurement of $\Sigma_i$ requires $O(k)$ effort,
measurement of an individual term of $V$ requires $O(q)$ effort, and the
evaluation of each bit of $p$ is repeated $m=O(\log(1/\epsilon_m))$ times
as in the previous subsection. The total measurement effort is, therefore,
$O(Kq^2 \cdot kq \cdot qm)=O(Kkq^4m)$.
To evaluate $\langle O\rangle$ with fractional error $\epsilon$, we need
$q=O(\log(1/\epsilon))$, as in Eq.(\ref{Oacc}). That makes the overall
measurement complexity for evaluation of $\langle O\rangle$
\begin{equation}
\label{cmplx_meas}
O(Kkq^4m) = O\left( Kk\log^4(1/\epsilon)\log(1/\epsilon_m) \right) ~.
\end{equation}
The spatial resources needed for the measurement process are one $(n+q)$-bit
register to hold the digital representation of $|x\rangle$, and several $q$-bit
registers that hold the $q^2$ measurement results (for each $\Sigma_i$) adding
up to $p$. The measurement process we have constructed is thus efficient,
belonging to the class P:P, i.e. polynomial in the input size and the output
accuracy.

We point out that the preceding strategy cannot be used to efficiently
evaluate $\|x\|^2 \equiv \langle x|I|x\rangle$, because it requires $O(N)$
contributions to be added together. For that reason, to carry out efficient
measurements, we need to restrict ourselves to measuring observables such
that $\|x\|$ is either known (e.g. unitary evolution) or cancels out (e.g.
ratios of expectation values).

\section{Efficient Solution of Linear Simultaneous Equations}
\label{effmatinv}

Now we turn to application of our measurement prescription to other linear
algebra problems. Consider the problem of solving $Ax=b$, where $A$ is an
$N \times N$ matrix and $x$,$b$ are $N$-component vectors. The formal
solution, $x=A^{-1}b$, involves matrix inversion. Exact computation of
the matrix inverse is time consuming, e.g. Gaussian elimination requires
$O(N^3)$ computational effort. So approximate iterative methods are
frequently used to solve this linear algebra problem, especially when the
matrix $A$ is sparse. They generate a sequence of approximate solutions,
until a termination criterion signals that convergence up to a specified
accuracy has been achieved. These methods generically work in the Krylov
space, ${\cal K}_r(A,x_0)={\rm span}\{x_0,Ax_0,A^2x_0,\ldots,A^{r-1}x_0\}$,
starting with an initial guess $x_0$. For a sparse $A$, each matrix-vector
multiplication is $O(N)$, and this space can be covered with $O(rN)$ effort.

The problem is singular when an eigenvalue of the matrix $A$ approaches
zero, and so the iterative convergence depends on the matrix condition
number $\kappa$ that is the ratio of the largest to the smallest eigenvalue
magnitudes. Within this context, the computational complexity of approximate
iterative inversion methods is characterised in terms of the matrix size $N$,
the desired solution accuracy $\epsilon$ and the matrix condition number
$\kappa$. For the popular conjugate gradient algorithm \cite{CGorig},
the computational complexity of solving $Ax=b$ on classical computers is
$O(dN\sqrt{\kappa}\log(1/\epsilon))$, for a positive definite and $d$-sparse
$A$.

In this Section, we present a simple quantum algorithm for computation of
$x=A^{-1}b$ based on the Newton-Raphson method. We assume that the matrix
$A$ has certain physically motivated properties, analogous to those of $H$
in the Hamiltonian evolution problem: (i) the Hilbert space is a tensor
product of many small components, (ii) the components have only local
interactions that make $A$ sparse, and (iii) both $A$ and $b$ are specified
in terms of a finite number of efficiently computable functions, so that
the resources needed to write them down do not influence the computational
complexity. On the other hand, we do not need to assume that $A$ is positive
definite. Then the algorithm is easy to implement with our digital
representation of quantum states, and has computational complexity
$O(dn\kappa^2\log^4(1/\epsilon))$.

\subsection{Solution by Newton-Raphson Method}

The Newton-Raphson method can be used to iteratively solve an algebraic
equation, when the derivatives of the functions involved can be easily
obtained. It can be applied to the matrix inversion problem, with the
number of significant digits approximately doubling with every iteration
\cite{NumRecip}.

Let $B_0$ be a suitable initial guess for $A^{-1}$. Then, in terms of the
residual matrix $R=I-B_0 A$, the partial sum
\begin{equation}
B_r = (1+R+R^2+\ldots+R^r) B_0
\end{equation}
converges to $A^{-1}$ as $n\rightarrow\infty$. The Newton-Raphson recurrence
relation,
\begin{equation}
B_{2r+1} = 2B_r - B_r A B_r ~,~~ {\rm r=0,1,3,7,15,\ldots} ~,
\end{equation}
converges quadratically to $A^{-1}$, doubling the order $r$ at each stage.
The iterative solution, $x_r=B_r b$, satisfies
\begin{equation}
\label{nriter}
x_{r+1} = x_r + B_0 (b - A x_r) ~,
\end{equation}
and converges geometrically to the desired solution:
\begin{eqnarray}
x_{r+1}-x &=& (I-B_0 A) (x_r-x) ~,\\
          &=& (I-B_0 A)^{r+1} (x_0-x) ~,\\
          &=& -(I-B_0 A)^{r+2} x ~.
\end{eqnarray}
Formally, $\|x_r - x\| \leq \|I-B_0 A\|^{r+1} \cdot \|x\|$
provides the fractional accuracy bound for $x$.
We can always scale the problem such that $\|A\|$ is $\Theta(1)$.
Then $\|A^{-1}\|=\Theta(\kappa)$, and $\|x\|=\Theta(\kappa\|b\|)$.

A good initial choice is $B_0=\alpha A^\dagger$, with
$\alpha=2/(\lambda_{\rm min}+\lambda_{\rm max})$ in terms of the eigenvalues
of $A^\dagger A$ \cite{PanReif}. The condition number of $A^\dagger A$ is
$\lambda_{\rm max}/\lambda_{\rm min}=\kappa^2$. That makes
\begin{equation}
\|I-B_0 A\| = \left| \frac{\lambda_{\rm max}-\lambda_{\rm min}}
                          {\lambda_{\rm max}+\lambda_{\rm min}} \right|
            = \frac{\kappa^2 - 1}{\kappa^2 + 1} ~.
\end{equation}
Each iteration of Eq.(\ref{nriter}) requires two matrix-vector products.
So the computational complexity of the algorithm, for reaching the accuracy
$\|\Delta x\| < \epsilon \|x\|$, is
\begin{equation}
\label{NRcomplexity}
2r{\cal C} = O\left( \frac{2\log(1/\epsilon)}{\log(1/\|I-B_0 A\|)}
                     {\cal C} \right)
           = O(\kappa^2 \log(1/\epsilon) {\cal C}) ~,
\end{equation}
where ${\cal C}$ is the computational cost of a matrix-vector product.
In practice, Frobenius bounds can be used to estimate the eigenvalue range
of $A^\dagger A$, and the choice
\begin{equation}
\alpha=\frac{1}{\sum_{jk}|A_{jk}|^2} \qquad{\rm or}\qquad
\alpha=\frac{1}{\left( \mathop{\max}\limits_j \sum_{k}|A_{jk}| \right)
                \left( \mathop{\max}\limits_k \sum_{j}|A_{jk}| \right)} ~,
\end{equation}
guarantees convergence of the algorithm \cite{PanReif}. With our assumption
that $A$ is specified in terms of a finite number of efficiently computable
functions, such a choice of $\alpha$ can be easily made at the start of the
algorithm. Note that this algorithm does not require the matrix $A$ to be
positive definite.

For sparse matrices, representing local interactions, calculation of the
matrix-vector product can be efficiently parallelised and carried out in
superposition on a quantum computer, as described in the next subsection.
Coding the $N$ components of $x$ using $n=\log_2 N$ qubits, we have
${\cal C}=O(dnq^3)$, where $d$ is the sparsity of the matrix and $q$ is
the precision (number of register bits) required for individual arithmetic
operations. As explained in Section \ref{digitalqc}, we can keep the
round-off errors under control by choosing
\begin{equation}
\label{NRqvalue}
q = \Omega\Big(\log\big(\frac{dr}{\epsilon\|x\|}\big)\Big)
  = \Omega\Big(\log\big(\frac{dr\|A\|}{\kappa\epsilon\|b\|}\big)\Big) ~.
\end{equation}
The net computational complexity of the algorithm is, therefore,
\begin{equation}
\label{cmplx_NR}
O\left( dn\kappa^2\log(\frac{1}{\epsilon})
        \log^3(\frac{d\kappa\log(1/\epsilon)\|A\|}{\epsilon\|b\|}) \right) ~.
\end{equation}

While the (poly)logarithmic dependence of the computational complexity on
the matrix size $N$ and the accuracy $\epsilon$ are desired features, the
quadratic dependence on the condition number $\kappa$ remains a hurdle to
be overcome. A known optimisation strategy is to use parallelisable
preconditioners to effectively reduce $\kappa$ (see for example,
Ref.~\cite{clader}). The existence and construction of such preconditioners,
however, depends on the detailed properties of the matrix to be inverted
(e.g. FFT for translationally invariant matrices).

For comparison, the classical conjugate gradient algorithm also converges
geometrically to the desired solution, but requires the matrix to be
inverted to be positive definite in order to guarantee convergence.
When $A$ is not positive definite, the problem solved in practice is
$A^\dagger Ax=A^\dagger b$, even at the cost of squaring the matrix
condition number. Its computational complexity is then
$O(\kappa\log(1/\epsilon){\cal C}_{CG})$ when $A$ is not positive definite,
and $O(\sqrt{\kappa}\log(1/\epsilon){\cal C}_{CG})$ when $A$ is positive
definite, where ${\cal C}_{CG}$ is the computational effort to implement
a single iteration of the algorithm. ${\cal C}_{CG}$ involves matrix-vector
multiplications, which can be easily parallelised for sparse matrices, and
calculation of inner products for evolving the solution vector in the Krylov
space along orthogonal directions, which cannot be parallelised. Overall,
${\cal C}_{CG} = O(dNq^3)$. Evolution along orthogonal directions makes the
conjugate gradient algorithm converge faster than our Newton-Raphson method
based algorithm, in terms of the dependence on $\kappa$. That is the price
paid for bypassing the calculation of inner products, in order to achieve
quantum superposition that reduces the computational complexity dependence
from $N$ to $n$.

Quantum algorithms that reduce the computational complexity dependence from
$N$ to $n$ have been proposed before; the algorithm of Ref.~\cite{harrow}
has computational complexity $O(d^2n\frac{\kappa^2}{\epsilon})$, and the
algorithm of Ref.~\cite{clader} improved the dependence on $\kappa$ using
a preconditioning matrix. Both these algorithms are based on Hamiltonian
simulation of $e^{iAt}$. The phase estimation technique is used to estimate
the eigenvalues of $A$ up to a specified level of precision, and then the
matrix is inverted in the spectral basis. Phase estimation has to evaluate
the eigenvalues with error $O(\frac{\epsilon}{\kappa})$, which contributes
the dominant factor $\frac{\kappa}{\epsilon}$ to the computational complexity.
Using a different strategy, we have achieved an exponential improvement in
the dependence of the computational complexity on $\epsilon$.

\subsection{Digital State Implementation}
\label{nrdigit}

Evaluation of $x_r$ by iterating Eq.(\ref{nriter}), starting with
$x_0=\alpha A^\dag b$, requires multiplication of a vector by a constant,
addition of two vectors, and multiplication of a matrix with a vector.
The first two operations are easily carried out with the digital
representation described in Section \ref{digitalqc}. Multiplication of a
matrix with a vector, on the other hand, has to be carefully implemented
such that quantum superposition converts its computational complexity from
classical $O(N)$ to quantum $O(n)$. 

A simple way to parallelise multiplication of the sparse matrix with
a vector is to decompose the matrix as a sum of block-diagonal parts,
$A=\sum_{i=1}^d A_i$, with each part consisting of a large number of
mutually independent blocks. Any sparse matrix can be efficiently
decomposed in this manner, according to Vizing's theorem, using an
edge-colouring algorithm for the corresponding graph. Then each colour
represents a part, and each part contains $O(N/2)$ mutually independent
$2\times2$ blocks.%
\footnote{In many physical problems, the matrix $A$ is Hermitian and
          the off-diagonal elements, $A_{j,j+\mu} = A^*_{j+\mu,j}$,
          can be denoted by a single edge of the graph. Otherwise,
          the edges of the graph have to be directed.}
Simultaneously carrying out individual block calculations for each part
with a superposition of their block labels, and evaluating the contribution
of each part in succession, the total computational effort for sparse
matrix-vector multiplication becomes $O(d\log(N/2))$ times the effort for
a single $2\times2$ block multiplication.

In the digital representation, multiplication by diagonal matrix elements
is straightforward, and multiplication by off-diagonal matrix elements
of the $2\times2$ blocks becomes straightforward provided one can swap
the $q$-bit register values, i.e.
\begin{equation}
\label{swapx}
|j\rangle |x_j\rangle + |j+\mu\rangle |x_{j+\mu}\rangle \longrightarrow
|j\rangle |x_{j+\mu}\rangle + |j+\mu\rangle |x_j\rangle ~.
\end{equation}
Such a swap operation is performed by the reflection operator,
\begin{equation}
S = \sigma_1 \otimes I^{\otimes q} ~,~~ S^2 = I ~,
\end{equation}
acting on the subspace
$\{|j\rangle,|j+\mu\rangle\} \otimes \{|x_j\rangle,|x_{j+\mu}\rangle\}$.
The swap can be undone after the off-diagonal matrix element multiplication
for a particular part $A_i$, to use $|x_j\rangle$ again for the next part.

\begin{figure}[t]
{
\begin{center}
%\setlength{\unitlength}{1mm}
% Standard unit length is 1pt
\begin{picture}(400,180)
\put(20,180){\line(1,0){160}}
\put(20,160){\line(1,0){55}}
\put(85,160){\line(1,0){50}}
\put(145,160){\line(1,0){35}}
\put(25,140){\line(1,0){155}}
\put(20,120){\line(1,0){160}}
\put(20,100){\line(1,0){160}}
\put(20,80){\line(1,0){15}}
\put(45,80){\line(1,0){135}}
\put(20,60){\line(1,0){30}}
\put(70,60){\line(1,0){110}}
\put(20,40){\line(1,0){90}}
\put(130,40){\line(1,0){50}}
\put(40,20){\line(1,0){55}}
\put(105,20){\line(1,0){50}}
\put(165,20){\line(1,0){15}}

\put(185,180){\ldots\ldots}
\put(185,160){\ldots\ldots}
\put(185,140){\ldots\ldots}
\put(185,120){\ldots\ldots}
\put(185,100){\ldots\ldots}
\put(185,80){\ldots\ldots}
\put(185,60){\ldots\ldots}
\put(185,40){\ldots\ldots}
\put(185,20){\ldots\ldots}

\put(0,178){$|0\rangle_q$}
\put(0,158){$|0\rangle_q$}
\put(0,138){$|b_j\rangle_q$}
\put(0,118){$|i\rangle$}
\put(0,98){$|j\rangle$}
\put(0,78){$|0\rangle$}
\put(0,58){$|0\rangle_q$}
\put(0,38){$|0\rangle_q$}
\put(0,18){$|(x_r)_j\rangle_q$}

\put(40,120){\circle*{4}}
\put(40,100){\circle*{4}}
\put(40,85){\line(0,1){35}}
\put(35,75){\framebox(10,10){$\mu$}}

\put(60,140){\circle*{4}}
\put(60,140){\line(0,1){25}}
\put(55,155){\framebox(10,10){}}

\put(60,120){\circle*{4}}
\put(60,100){\circle*{4}}
\put(60,65){\line(0,1){55}}
\put(50,55){\framebox(20,10){$-A$}}

\put(80,20){\circle*{4}}
\put(80,60){\circle*{4}}
\put(80,20){\line(0,1){135}}
\put(75,155){\framebox(10,10){{\huge$\times$}}}

\put(100,100){\circle*{4}}
\put(100,80){\circle*{4}}
\put(100,25){\line(0,1){75}}
\put(95,15){\framebox(10,10){$S$}}

\put(120,120){\circle*{4}}
\put(120,100){\circle*{4}}
\put(120,80){\circle*{4}}
\put(120,45){\line(0,1){75}}
\put(110,35){\framebox(20,10){$-A$}}

\put(140,20){\circle*{4}}
\put(140,40){\circle*{4}}
\put(140,20){\line(0,1){135}}
\put(135,155){\framebox(10,10){{\huge$\times$}}}

\put(160,100){\circle*{4}}
\put(160,80){\circle*{4}}
\put(160,25){\line(0,1){75}}
\put(155,15){\framebox(10,10){$S$}}

\put(107,10){$|(x_r)_{j+\mu_i}\rangle$}
\put(170,165){$|(b-Ax_r)_j\rangle_q$}
\put(175,50){$|(-A_i)_{j,j}\rangle$}
\put(170,30){$|(-A_i)_{j,j+\mu_i}\rangle$}
\put(180,10){$|(x_r)_j\rangle_q$}

\put(215,180){\line(1,0){50}}
\put(275,180){\line(1,0){50}}
\put(335,180){\line(1,0){35}}
\put(215,160){\line(1,0){70}}
\put(295,160){\line(1,0){50}}
\put(355,160){\line(1,0){15}}
\put(215,140){\line(1,0){155}}
\put(215,120){\line(1,0){155}}
\put(215,100){\line(1,0){155}}
\put(215,80){\line(1,0){155}}
\put(215,60){\line(1,0){25}}
\put(260,60){\line(1,0){110}}
\put(215,40){\line(1,0){85}}
\put(320,40){\line(1,0){50}}
\put(215,20){\line(1,0){155}}

\put(230,20){\circle*{4}}
\put(230,20){\line(0,1){165}}
\put(225,175){\framebox(10,10){}}

\put(250,120){\circle*{4}}
\put(250,100){\circle*{4}}
\put(250,65){\line(0,1){55}}
\put(240,55){\framebox(20,10){$\alpha A^\dag$}}

\put(270,60){\circle*{4}}
\put(270,160){\circle*{4}}
\put(270,60){\line(0,1){115}}
\put(265,175){\framebox(10,10){{\huge$\times$}}}

\put(290,80){\circle*{4}}
\put(290,100){\circle*{4}}
\put(290,80){\line(0,1){75}}
\put(285,155){\framebox(10,10){$S$}}

\put(310,120){\circle*{4}}
\put(310,100){\circle*{4}}
\put(310,80){\circle*{4}}
\put(310,45){\line(0,1){75}}
\put(300,35){\framebox(20,10){$\alpha A^\dag$}}

\put(330,40){\circle*{4}}
\put(330,160){\circle*{4}}
\put(330,40){\line(0,1){135}}
\put(325,175){\framebox(10,10){{\huge$\times$}}}

\put(350,80){\circle*{4}}
\put(350,100){\circle*{4}}
\put(350,80){\line(0,1){75}}
\put(345,155){\framebox(10,10){$S$}}

\put(375,178){$|(x_{r+1})_j\rangle_q$}
\put(375,78){$|j+\mu_i\rangle$}
\put(375,58){$|(\alpha A^\dag_i)_{j,j}\rangle$}
\put(375,38){$|(\alpha A^\dag_i)_{j,j+\mu_i}\rangle$}

\end{picture}
\end{center}
}
\caption{Digital quantum logic circuit for implementing the recursion
relation, Eq.(\ref{nriter}), to be executed with a uniform superposition
over the index $j$. Operations for a single $A_i$ containing only $2\times2$
blocks (labeled by $j,j+\mu_i$) are shown. The two sparse matrix-vector
multiplications involved in Eq.(\ref{nriter}) are separated by ``\ldots",
and a sum over the index $i$ is implicit in each of these two parts. Among
the controlled logic gates, \frame{\strut$~\mu~$}, \frame{\strut$~-A~$} and
\frame{\strut$~\alpha A^\dag~$} denote oracle operations specified by the
matrix, \frame{\strut$~S~$} is the swap operation of Eq.(\ref{swapx}),
\frame{\large$\times$} stands for the generalised Toffoli gate implementing
$|a,b,c\rangle \rightarrow |a,b,c+ab\rangle$, and \frame{\large$+$} labels
generalised C-not gates performing $|a,b\rangle \rightarrow |a,b+a\rangle$.}
\label{fragnriter}
\end{figure}
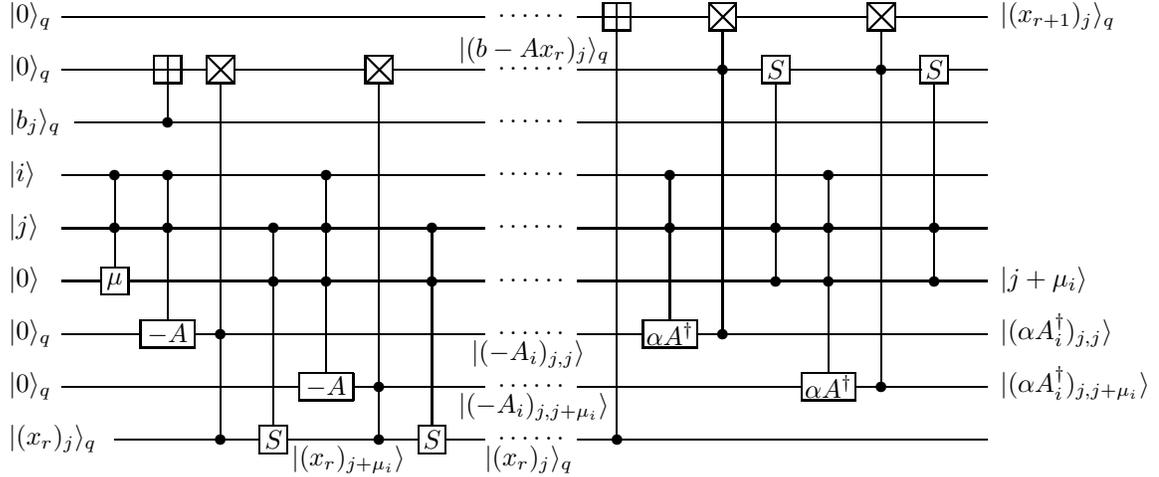

The digital circuit implementation of Eq.(\ref{nriter}), is schematically
illustrated in Fig.\ref{fragnriter}. For the sake of clarity, only the
steps corresponding to a single $A_i$ are shown, with the sum over $i$
left implicit. The oracles have computational complexity $O(q^3)$ arising
from evaluation of the matrix elements; the rest of the linear algebra
operations have computational complexity $O(q^2)$. Including contributions
of all the parts, and the computational effort needed to superpose the
index $j$, we thus have the time complexity ${\cal C}=O(dnq^3)$ per
iteration. We also note that the space resources required to put together
the final solution $|x\rangle$ are a fixed number of $n$-bit and $q$-bit
registers, because these registers can be reused.

We repeat that in our construction based on the digital representation for
the quantum states, the full quantum advantage that reduces $N$ to $n$ in
the computational complexity arises from a simple superposition of the
quantum state label $j$, and this superposition in turn requires
decomposition of the matrix into block-diagonal parts.

Once a sufficiently accurate solution for $|x\rangle=A^{-1}|b\rangle$
is obtained in a quantum register, the methodology of the Section
\ref{effmeas} can be used to efficiently evaluate $k$-local observables
$\langle x|O|x\rangle$. 

\section{Efficient Matrix Exponentiation}
\label{effmatexp}

Efficient exponentiation of a Hermitian matrix $A$ is another algorithm that
we construct, assuming that $A$ satisfies the three properties listed in the
beginning of Section \ref{effmatinv}. We restrict ourselves to evaluation of
$e^{-At}|b\rangle$, for given $t$ and $|b\rangle$, which is the combination
often encountered in physical applications.

The eigenvalue spectrum of any bounded Hermitian matrix is within a range
$[\lambda_{\rm min},\lambda_{\rm max}]$. With a linear transformation,
this range can be mapped to any desired real interval, and the additive
and the multiplicative constants can be handled by simple rescaling.
For example, 
\begin{equation}
\tilde{A} = (A-\lambda_{\rm min}I) / (\lambda_{\rm max}-\lambda_{\rm min}) ,
\end{equation}
makes $\tilde{A}$ positive semidefinite, with eigenvalues in $[0,1]$.
Then $e^{-At}=e^{-\lambda_{\rm min}t} e^{-\tilde{A}\tilde{t}}$, with
\begin{equation}
\tilde{t} = t (\lambda_{\rm max}-\lambda_{\rm min}) ~.
\end{equation}
In situations where $\lambda_{\rm min}$ and $\lambda_{\rm max}$ are not
exactly known, respectively lower and upper bounds for them can be used.
Frobenius bounds are convenient, based on the constraint that every
eigenvalue of $A$ lies in one of the disks centred at $A_{jj}$ with radius
$\sum_{l\ne j} |A_{lj}|$, and can be easily obtained for an $A$ that is
specified in terms of a finite number of efficiently computable functions.
Henceforth, we assume that such a rescaling of $A$ has been carried out as
per the need of the algorithm, and drop the tilde's on $A$ and $t$ for
simplicity.

\subsection{Chebyshev Expansion and its Complexity}
\label{expcheby}

Chebyshev polynomials provide uniform approximations for bounded functions,
with fast convergence of the series \cite{Arfken,TalEzer}. We scale $A$ such
that its eigenvalue spectrum is within the domain $[-1,1]$ of the Chebyshev
polynomials $T_m(x)=\cos(m\cos^{-1}x)$. Then $\|e^{-At}\|=\Theta(e^t)$, and
\begin{equation}
\label{chebyexp}
e^{-At} = \sum_{k=0}^\infty C_k(t)~T_k(A) ~,
\end{equation}
where the expansion coefficients are the Bessel functions:
\begin{eqnarray}
C_0 &=& \frac{1}{\pi} \int_0^\pi e^{-t\cos\theta} d\theta = I_0(t) ~, \\
C_{k>0} &=& \frac{2}{\pi} \int_0^\pi e^{-t\cos\theta} \cos(k\theta) ~d\theta 
     = 2(-1)^k I_k(t) ~.
\end{eqnarray}
Note that the Chebyshev polynomials are bounded in their domain, and the
coefficients $I_k(t) = t^k/(2^k k!)+\ldots$ fall off faster by a factor
of $2^k$ compared to the corresponding coefficients $t^k/k!$ of the
Taylor series expansion. This is the well-known advantage of the Chebyshev
expansion compared to other series expansions.

The modified Bessel functions obey $I_k(t)>0$ for $t>0$, and increase with
$t$ monotonically. Furthermore,
\begin{equation}
\label{Bessfn}
I_k(t) = \sum_{s=0}^\infty \frac{(t/2)^{k+2s}}{s!(k+s)!}
       < \frac{(t/2)^k}{k!} \exp\left(\frac{t^2}{4(k+1)}\right) ~.
\end{equation}
Therefore, when the Chebyshev expansion in Eq.(\ref{chebyexp}) is truncated
at order $r$, the truncation error is bounded by
\begin{eqnarray}
\sum_{k=r+1}^\infty 2I_k(t) &<& \sum_{k=r+1}^\infty
   2\frac{(t/2)^k}{k!} e^{t^2 / 4(k+1)} \\
   &<& 2\frac{(t/2)^{r+1}}{(r+1)!} e^{t^2 / 4(r+2)}
   \left(1 - \frac{t}{2(r+2)}\right)^{-1} . \nonumber
\end{eqnarray}
To control this truncation error, we choose
\begin{equation}
\label{rbound1}
r+2 \ge t ~,
\end{equation}
and then demand
\begin{equation}
\label{Ikbound}
\sum_{k=r+1}^\infty 2I_k(t) <
   4\frac{(t/2)^{r+1}}{(r+1)!} e^{t^2 / 4(r+2)} < \epsilon_0 ~.
\end{equation}
The formal solution, consistent with Eq.(\ref{rbound1}), is
\begin{equation}
\label{chebyiter}
r = \frac{e^{5/4}t}{2} + {\rm ln}(1/\epsilon_0)
  = O(t+\log(1/\epsilon_0)) ~,
\end{equation}
and the fractional accuracy of this truncation is $\epsilon=e^{-t}\epsilon_0$.

A truncated series of the Chebyshev expansion is efficiently evaluated
using Clenshaw's algorithm, based on the recursion relation
\begin{equation}
T_{k+1}(A) = 2A~T_k(A) - T_{k-1}(A) ~,
\end{equation}
To evaluate $e^{-At}|b\rangle$, one initialises the vectors
$|y_{r+1}\rangle = 0$, $|y_r\rangle = C_r|b\rangle$,
and then uses the reverse recursion
\begin{equation}
\label{clenshawrec}
|y_k\rangle = C_k|b\rangle + 2A~|y_{k+1}\rangle - |y_{k+2}\rangle ~,
\end{equation}
from $k=r-1$ to $k=0$. At the end,
\begin{equation}
\sum_{k=1}^r C_k T_k(A) |b\rangle =
(C_0|b\rangle + |y_0\rangle - |y_2\rangle)/2
\end{equation}
is obtained using $r$ sparse matrix-vector products involving $A$.
The computational complexity of the procedure is then
\begin{equation}
\label{Chebycomplexity}
O(r{\cal C}_C) = O\left( (t+\log(1/\epsilon_0)) {\cal C}_C \right) ~,
\end{equation}
where ${\cal C}_C$ is the computational cost of implementing the recursion
of Eq.(\ref{clenshawrec}). As per the digital state implementation described
in the next subsection, ${\cal C}_C = O(dnq^3)$.

The Bessel functions $I_k(t)$ up to order $r$ can be efficiently calculated
to high precision, using the recursion relation
\begin{equation}
I_{k-1}(t) = \frac{2k}{t} I_k(t) + I_{k+1}(t) ~,
\end{equation}
in descending order \cite{AMS55}. One starts with approximate guesses for
$I_l(t)$ and $I_{l+1}(t)$, with $l$ slightly larger than $r$, and uses the
recursion relation repeatedly to reach $I_0(t)$. Then all the values are
scaled to the correct normalisation by imposing the constraint
$I_0(t) + 2\sum_{k=1}^{\lceil l/2 \rceil} I_{2k}(t) = e^t$. This procedure
to determine the expansion coefficients requires $\Theta(rq^2)$ computational
effort, and so does not alter the overall computational complexity given by
Eq.(\ref{Chebycomplexity}).

To control the round-off error while summing up the Chebyshev expansion,
the coefficients $C_k$ up to order $r$, and the elements of matrix $A$,
have to be evaluated to $q = \Omega(\log(\frac{dr}{\epsilon}))$
bit precision, as explained in Section \ref{digitalqc}.%
\footnote{Note that $q$ is determined by the fractional accuracy of the
intended result.}
The net computational complexity of the algorithm is, therefore,
\begin{equation}
\label{cmplx_exp}
O\left( dn\Big(t+\log(\frac{1}{e^t\epsilon})\Big)
        \log^3(\frac{d(t+\log(1/e^t\epsilon))}{\epsilon}) \right) ~.
\end{equation}
The desirable features of this computational complexity are the linear
dependence on $n$ and $t$, and the (poly)logarithmic dependence on $\epsilon$.
This behaviour is similar to the scaling in Eq.(\ref{cmplx_hamevol}) for
the Hamiltonian evolution problem.

\subsection{Digital State Implementation}
\label{chebyser}

\begin{figure}[t]
{
\begin{center}
%\setlength{\unitlength}{1mm}
% Standard unit length is 1pt
\begin{picture}(250,180)
\put(20,180){\line(1,0){35}}
\put(65,180){\line(1,0){10}}
\put(85,180){\line(1,0){50}}
\put(145,180){\line(1,0){30}}
\put(185,180){\line(1,0){5}}
\put(25,160){\line(1,0){165}}
\put(25,140){\line(1,0){165}}
\put(20,120){\line(1,0){170}}
\put(20,100){\line(1,0){170}}
\put(20,80){\line(1,0){15}}
\put(45,80){\line(1,0){145}}
\put(20,60){\line(1,0){35}}
\put(65,60){\line(1,0){125}}
\put(20,40){\line(1,0){95}}
\put(125,40){\line(1,0){60}}
\put(40,20){\line(1,0){55}}
\put(105,20){\line(1,0){50}}
\put(165,20){\line(1,0){25}}
\put(40,0){\line(1,0){150}}

\put(0,178){$|0\rangle_q$}
\put(0,158){$|C_k\rangle_q$}
\put(0,138){$|b_j\rangle_q$}
\put(0,118){$|i\rangle$}
\put(0,98){$|j\rangle$}
\put(0,78){$|0\rangle$}
\put(0,58){$|0\rangle_q$}
\put(0,38){$|0\rangle_q$}
\put(0,18){$|(y_{k+1})_j\rangle_q$}
\put(0,-2){$|(y_{k+2})_j\rangle_q$}

\put(40,120){\circle*{4}}
\put(40,100){\circle*{4}}
\put(40,85){\line(0,1){35}}
\put(35,75){\framebox(10,10){$\mu$}}

\put(60,160){\circle*{4}}
\put(60,140){\circle*{4}}
\put(60,140){\line(0,1){35}}
\put(55,175){\framebox(10,10){{\huge$\times$}}}

\put(60,120){\circle*{4}}
\put(60,100){\circle*{4}}
\put(60,65){\line(0,1){55}}
\put(55,55){\framebox(10,10){$A$}}

\put(80,20){\circle*{4}}
\put(80,60){\circle*{4}}
\put(80,20){\line(0,1){155}}
\put(75,175){\framebox(10,10){{\huge$\times$}}}

\put(100,100){\circle*{4}}
\put(100,80){\circle*{4}}
\put(100,25){\line(0,1){75}}
\put(95,15){\framebox(10,10){$S$}}

\put(120,120){\circle*{4}}
\put(120,100){\circle*{4}}
\put(120,80){\circle*{4}}
\put(120,45){\line(0,1){75}}
\put(115,35){\framebox(10,10){$A$}}

\put(140,20){\circle*{4}}
\put(140,40){\circle*{4}}
\put(140,20){\line(0,1){155}}
\put(135,175){\framebox(10,10){{\huge$\times$}}}

\put(160,100){\circle*{4}}
\put(160,80){\circle*{4}}
\put(160,25){\line(0,1){75}}
\put(155,15){\framebox(10,10){$S$}}

\put(180,0){\circle*{4}}
\put(180,0){\line(0,1){175}}
\put(175,175){\framebox(10,10){\large$-$}}

\put(107,10){$|(y_{k+1})_{j+\mu_i}\rangle$}
\put(195,178){$|(y_k)_j\rangle_q$}
\put(195,78){$|j+\mu_i\rangle$}
\put(195,58){$|(2A_i)_{j,j}\rangle$}
\put(190,38){$|(2A_i)_{j,j+\mu_i}\rangle$}
\put(195,18){$|(y_{k+1})_j\rangle_q$}

\end{picture}
\end{center}
}
\caption{Digital quantum logic circuit for executing the recursion relation
of Clenshaw's algorithm, Eq.(\ref{clenshawrec}), to be executed with a
uniform superposition over the index $j$. Operations for a single $A_i$
containing only $2\times2$ blocks (labeled by $j,j+\mu_i$) are shown.
Among the controlled logic gates, \frame{\strut$~\mu~$} and
\frame{\strut$~A~$} denote oracle operations specified by the matrix,
\frame{\strut$~S~$} is the swap operation of Eq.(\ref{swapx}), 
\frame{\large$\times$} stands for the generalised Toffoli gate
implementing $|a,b,c\rangle \rightarrow |a,b,c+ab\rangle$, and
\frame{\large$-$} labels the generalised C-not gate performing
$|a,b\rangle \rightarrow |a,b-a\rangle$.}
\label{fragcheby}
\end{figure}
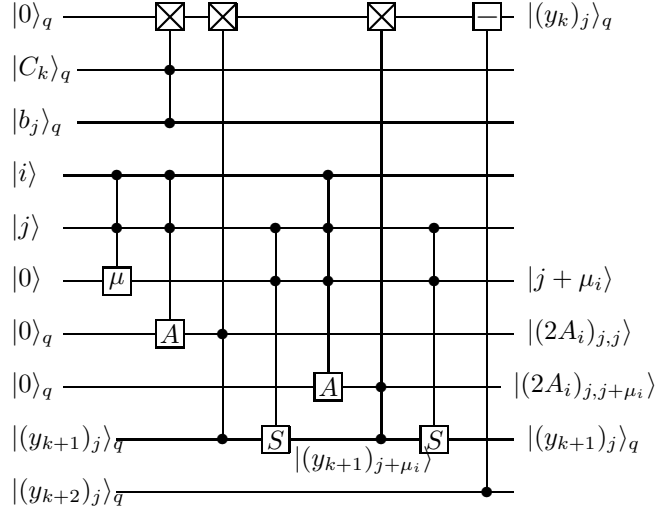

Summation of the series in Eq.(\ref{chebyexp}), truncated to order
$r$, requires $r$ executions of the Clenshaw recursion relation,
Eq.(\ref{clenshawrec}). For a sparse Hermitian $A$, that can be
implemented in the same manner as described in Section \ref{nrdigit}.
The digital circuit implementation of Eq.(\ref{clenshawrec}), for a
single part $A_i$, is schematically illustrated in Fig.\ref{fragcheby}.
It has computational complexity $O(q^3)$ arising from evaluation of the
matrix elements; the rest of the linear algebra operations contribute
computational complexity $O(q^2)$. Including contributions of all the
parts, and the computational effort needed to superpose the index $j$,
we thus have the time complexity ${\cal C}_C=O(dnq^3)$. We also note
that the space resources required to put together the full Chebyshev
expansion are a fixed number of $n$-bit registers and $O(r)$ $q$-bit
registers.

\subsection{Reduction of Matrix Inverse to Matrix Exponentiation}

For a positive definite matrix, the matrix inverse problem with accuracy
$\epsilon$ can be reduced to the matrix exponentiation problem, using the
result \cite{sachvish}
\begin{equation}
\label{matinvred}
(1-\epsilon)a^{-1} \le \sum_{j=p}^{p'} h~e^{jh}~e^{-ae^{jh}}
                   \le (1+\epsilon)a^{-1} ~,
\end{equation}
for all $a\in[\frac{1}{\kappa},1]$. Here $\epsilon$ is chosen as the
fractional accuracy for $a^{-1}$, since $a^{-1}$ is singular when
$\kappa\rightarrow\infty$. This result follows from approximating the
integral
\begin{equation}
a^{-1} = \int_0^\infty e^{-at} dt 
  \quad\mathop{=}\limits^{t=e^y}\quad \int_{-\infty}^\infty e^{y-ae^y} dy ~,
\end{equation}
by the trapezoidal rule, and bounding the error using the Euler-Maclaurin
formula. The choice of the discretisation parameters as \cite{sachvish}
\begin{equation}
h = \frac{2\pi}{e^2(2n+1)^2} ~,~~
n = \left\lceil \frac{1}{2}{\rm ln}\frac{24}{\epsilon} \right\rceil ~,
\end{equation}
\begin{equation}
p = \left\lfloor -\frac{1}{h}{\rm ln}\frac{3}{\epsilon} \right\rfloor ~,~~
p' = \left\lceil \frac{1}{h}{\rm ln}\left(
    \kappa~{\rm ln}\frac{3}{\epsilon}\right) \right\rceil ~,
\end{equation}
makes the computational effort needed for the reduction in
Eq.(\ref{matinvred}) polynomial in $\log(1/\epsilon)$ and $\log(\kappa)$.
Note that for the variable $t=e^{jh}$, we have
\begin{equation}
j\in[p,p'] ~\Longrightarrow~ t\in[\epsilon/3,\kappa~{\rm ln}(3/\epsilon)] ~.
\end{equation}
The result of Eq.(\ref{matinvred}) can be applied to positive definite
matrices, with $a \rightarrow A$, interpreting all matrix functions as
their power series expansions. Then rapidly converging Chebyshev expansion
for $\exp(-At)$ can be used, as described in Section \ref{expcheby}, for
every value of $j$.

The trapezoidal rule sum of Eq.(\ref{matinvred}) has $p'-p+1$ terms, and
its error has to be bounded by $\epsilon_0\le\kappa\epsilon$. That can be
achieved by choosing the Chebyshev expansion truncation order, as per
Eq.(\ref{chebyiter}),
\begin{equation}
r_j = O\Big(t+\log\big(\frac{p'-p}{\epsilon_0}\big)\Big)
    = O\Big(e^{jh} + \log\big(\frac{\log((\kappa/\epsilon)\log(1/\epsilon))}
                                   {\kappa\epsilon}\big)\Big) ~.
\end{equation}
The total number of times the recursion relation of Eq.(\ref{clenshawrec})
needs to be executed is then
\begin{eqnarray}
r_{\rm tot} = \sum_{j=p}^{p'} r_j
           &=& O\Big(\frac{\kappa}{h}\log\frac{1}{\epsilon}
  + \frac{1}{h}\log\big(\frac{\kappa}{\epsilon}\log\frac{1}{\epsilon}\big)
    \log\big(\frac{\log((\kappa/\epsilon)\log(1/\epsilon))}
                  {\kappa\epsilon}\big)\Big) \nonumber\\
           &=& O\big(\kappa\log^3(1/\epsilon) + \log^4(1/\epsilon)\big) ~.
\end{eqnarray}

For any fixed $b$, the fractional error in evaluating $x=A^{-1}b$ is the
same as the fractional error in evaluating $A^{-1}$. The preceding algorithm
has two sources of error: replacement of the integral by the Euler-Maclaurin
formula and truncation of the Chebyshev expansion. Putting all the pieces
together, the computational complexity of this algorithm for solving the
set of linear equations with accuracy $\|\Delta x\| < 2\epsilon \|x\|$,
is $O(r_{\rm tot}{\cal C}_C)$. Also, the required digital precision $q$ is
the same as in Eq.(\ref{NRqvalue}), with $r$ replaced by $r_{\rm tot}$.
Compared to Eq.(\ref{NRcomplexity}), we see that the dependence of the
computational complexity on $\log(1/\epsilon)$ is worse, although the
dependence on $\kappa$ is better when $A$ is positive definite. Overall,
the simplicity of implementation clearly favours the algorithm of Section
\ref{effmatinv}, for solution of linear simultaneous equations.

\subsection{Potential Applications}

With our digital representation, we can easily construct useful algorithms
involving unnormalised quantum states. Consider determination of expectation
values of various physical quantities in the ground state of a quantum system.
For Hamiltonian systems with a spectral gap $\Delta$, a convenient way to
obtain the ground state $|\psi_0\rangle$ is to evolve an approximate ansatz
for the ground state $|\psi\rangle$ in Euclidean time:
\begin{equation}
e^{-HT}|\psi\rangle = e^{-E_0 T} |\psi_0\rangle (1+O(e^{-\Delta T})) ~.
\end{equation}
The l.h.s. can be efficiently calculated by the methods presented earlier
in this Section, and then ratios of ground state expectation values can
be obtained even when the ground state energy $E_0$ is not known.

Going further, problems in statistical mechanics frequently involve systems
in equilibrium with a heat bath. Such a system with the Hamiltonian $H$ is
described by the thermal state
\begin{equation}
\rho = e^{-\beta H}/Z ~,~~ Z = Tr(e^{-\beta H}) ~.
\end{equation}
Physically observable quantities are then obtained as the expectation values
\begin{equation}
\langle O_a \rangle = Tr(\rho O_a) ~.
\end{equation}
For bounded Hamiltonians, $e^{-\beta H}$ is completely well-behaved,
and an expansion of $\rho$ in powers of $\beta$ has a non-zero radius
of convergence around $\beta=0$ (i.e. infinite temperature). Singular
critical phenomena arise from the large degeneracy of states that
contribute to the partition function $Z$. Our techniques, described
earlier in this Section, allow efficient calculation of $e^{-\beta H}$
acting on a vector. So by decomposing $O_a$ in the Pauli operator basis
(as described in Section \ref{effmeas}), we can efficiently evaluate
ratios of thermal state expectation values (where $Z$ cancels out).

\section{Summary and Outlook}

The quantum Hilbert space allows superposition of $N=2^n$ independent
components using $n$ qubits. They can then all be simultaneously processed
in the single-instruction-multiple-data mode, familiar from the design of
parallelisable algorithms for classical computers. This is a key ingredient
in development of quantum algorithms that can be exponentially faster than
their classical counterparts. Such a conversion of parallelisation into
superposition is not possible for generic computational problems, but it
can be achieved for many linear algebra problems, by domain decomposition
of the algorithms and breakdown of matrix operations into block-diagonal
ones.

On the other hand, this exponential advantage of quantum superposition is
severely limited by the fact that only $n$ bit worth of information can be
extracted from the result at the end. So the overall algorithm is efficient
only when the final observables are local in some sense. Although no general
prescription is available, we have shown that $k$-local observables appearing
in evaluations of $k$-point Green's functions of quantum many body problems
can be efficiently evaluated.

An important component of our demonstration is the digital representation
of quantum states \cite{qhamevol}. It makes linear algebra calculations
straightforward to perform. But, much more importantly, it also allows a
bit-by-bit deterministic evaluation of the expectation values, instead of
a probabilistic one. The well-known Chernoff bound strategy can then be
used to make the measurement effort logarithmic with respect to the output
accuracy. Thus, parallelisable algorithms with $k$-local observables become
efficient with respect to both the input and output sizes, and the problems
they solve belong to the computational complexity class that we have labeled
P:P.

Our quantum measurement strategy can be applied to any output state of a
quantum algorithm, as in the Hamiltonian evolution problem. But a useful
feature of the digital representation is that it is not constrained by
unitary evolution at every step, and so one can easily incorporate in it
non-unitary evolution steps such as series expansions. We have used this
flexibility to construct efficient quantum algorithms for two non-unitary
but practical problems: solution of simultaneous linear equations using the
Newton-Raphson method, and exponentiation of a matrix using the Chebyshev
expansion. It would certainly be worthwhile to explore other problems that
can be solved efficiently using our methods.

Our algorithm construction is explicit, and not reductionist, which clearly
demonstrates how the algorithms work in practice. It also illustrates how
the decomposition of matrices into block-diagonal components and the digital
representation of quantum states are closely tied, respectively, to the
input and the output efficiency of our algorithms. The classical ingredients
used in our algorithms---parallelisation of linear algebra operations,
digital representation, Chernoff bound, Newton-Raphson method, Chebyshev
expansion---are all well-known. Our contribution has been to put them
together in a manner that successfully carries over their advantages into
the quantum domain.

\end{document}